\documentclass[pra,twocolumn]{revtex4}
\usepackage{eurosym}
\usepackage{amssymb}
\usepackage{amsfonts}
\usepackage{amsmath}
\usepackage{graphicx}
\usepackage{bm}
\usepackage{braket}
\usepackage{mathtools}
\usepackage{textcomp}
\usepackage{xcolor}
\usepackage{float}

\begin{document}

\title{Vortex solitons in quasi-phase-matched photonic crystals with the
third harmonic generation}
\author{Xuening Wang$^{1}$}
\author{Yuxin Guo$^{1}$}
\author{Qiuyi Ning$^{1,2}$}
\author{Bin Liu$^{1,2}$}
\author{Hexiang He$^{1,2}$}
\email{sysuhhx@163.com}
\author{Li Zhang$^{1,2}$}
\email{zhangli4102@126.com}
\author{Boris A. Malomed$^{3,4}$}
\author{Yongyao Li$^{1,2}$}
\affiliation{$^1$School of Physics and Optoelectronic Engineering, Foshan University,
Foshan 528225, China\\
$^2$Guangdong-Hong Kong-Macao Joint Laboratory for Intelligent Micro-Nano
Optoelectronic Technology, Foshan University, Foshan 528225, China\\
$^3$Department of Physical Electronics, School of Electrical Engineering,
Faculty of Engineering, Tel Aviv University, Tel Aviv 69978, Israel\\
$^4$Instituto de Alta Investigaci\'{o}n, Universidad de Tarapac\'{a},
Casilla 7D, Arica, Chile }

\begin{abstract}
We report stable composite vortex solitons in the model of a
three-dimensional photonic crystal with the third-harmonic (TH) generation
provided by the quasi-phase-matched quadratic nonlinearity. The photonic
crystal is designed with a checkerboard structure in the $\left( x\text{,}%
y\right) $ plane, while the second-order nonlinear susceptibility, $d(z)$,
is modulated along the propagation direction as a chains of rectangles with
two different periods. This structure can be fabricated by means of
available technologies. The composite vortex solitons are built of
fundamental-frequency (FF), second-harmonic (SH), and TH components,
exhibiting spatial patterns which correspond to vortex  with
topological charges $s=1$, a quadrupole with $s=2$,  and an anti-vortex structure with $s = -1$,
respectively. The soliton
profiles feature rhombic or square patterns, corresponding to phase-matching
conditions $\varphi =0$ or $\pi $, respectively, the rhombic solitons
possessing a broader stability region. From the perspective of the
experimental feasibility, we show that both the rhombic and square-shaped
composite vortex solitons may readily propagate in the photonic crystals
over distances up to $\sim 1$ m. The TH component of the soliton with $s=\mp
1$ is produced by the cascaded nonlinear interactions, starting from the FF
vortex component with $s=\pm 1$ and proceeding through the quadrupole SH one
with $s=2$. These findings offer a novel approach for the creation and
control of stable vortex solitons in nonlinear optics.
\end{abstract}

\maketitle


\section{Introduction}

The stability of vortex solitons is a major topic in the field of nonlinear
optics~\cite{baip2022}. As optical fields carrying orbital angular momentum,
vortex beams offer applications to optical tweezers and particle
manipulation, super-resolution imaging, optical information processing, and
quantum communications~\cite{zcpb2012,wlpr2016,zsr2014,jnp2012}. However, in
media with the quadratic ($\chi ^{(2)}$) nonlinearity, vortex solitons are
generally unstable~\cite{apr2002,wprl1995,wol1995}. They are primarily
vulnerable to the azimuthal instability, which breaks the vortex soliton in
fragments, in the form of fundamental solitons~\cite%
{wprl1997,lel1997,dol1998}. The proneness to the fragmentation severely
limits potential applications of vortex solitons. Various stabilization
mechanisms have been proposed to mitigate the problem. In particular, the
use of competing nonlinearities provides an effective solution~\cite%
{ipre2001,dpre2004,pprl2000,dcsf2025,xoe2023,rol1992,cprl1995,soc1996,aol1995}%
. In cubic ($\chi ^{(3)}$) media, the introduction of higher-order
nonlinearities, such as quintic or logarithmic, may stabilize the solitons
against the self-focusing collapse and fragmentation~\cite{dpre2000}. In $%
\chi ^{(2)}$ media, the cascading mechanism may induce an effective
third-order nonlinearity $\chi _{\text{eff}}^{(3)}$, which exhibits a
self-defocusing effect under specific conditions, thereby stabilizing vortex
solitons~\cite{lol1994,cpio2000}. However, the\ realization of the cascading
mechanism in $\chi ^{(2)}$ materials requires an optical intensity $\sim 10~%
\text{GW/cm}^{2}$ to effectively emulate $\chi ^{(3)}$ nonlinear effects. So
high light intensities are often difficult to achieve, and may be close to
the damage threshold of nonlinear crystals.

Recently, other stabilization mechanisms for vortex solitons have been
explored \cite%
{ccsf2024,wpla2025,bnp2019,bp2021,hpra2022,hpra2020,cprl2024,jcsf2023,qol2021}%
. In this vein, the stable propagation of vortex solitons was predicted in
three-dimensional quasi-phase-matched (QPM) $\chi ^{(2)}$ photonic crystals
\cite{fprl2023}. This study relied upon a checkerboard structure, where
periodic modulation of the $\chi ^{(2)}$ coefficient leads to the formation
of stable vortex solitons, enhancing the nonlinear coupling between their
fundamental-frequency (FF) and the second-harmonic (SH) components However,
the study dealt solely with the SH generation, and did not address the
third-harmonic generation (THG) and its impact on the vortex-soliton
stability. Efficient THG in a $\chi ^{(2)}$ nonlinear crystal, including
enhanced nonlinear coupling between the FF, SH and TH (third-harmonic)
components, provided by a properly designed QPM structure, was demonstrated
in Ref.~\cite{ss1997}. However, the primary focus of that work was on
improving the harmonic-conversion efficiency, rather than stability of
self-trapped beams. In fact, the propagation of vortex solitons in the
presence of THG remains largely unexplored.

The present work aims to propose a scheme for the stabilization of vortex
solitons in a $\chi ^{(2)}$ medium. To this end, we design an appropriate
QPM structure in the form of a checkerboard-polarized pattern~\cite%
{zpre2025,jcsf2024,soe2024,ycpl2024}, which can be realized using
femtosecond laser engineering techniques~\cite%
{tnp2018,dnp2018,snp2018,alsa2021,alpr2010,hfo2020,saom2023}. Unlike
conventional periodically poled structures, the checkerboard one exhibits
two-dimensional periodicity in the transverse $(X,Y)$ plane, enabling
multidirectional nonlinear coupling, rather than being limited to a
one-dimensional QPM modulation~\cite{mjpd1995,tol2000,aapl2003}. The
two-dimensional modulation provides improved conditions for the formation
and stable propagation of composite vortex solitons. Furthermore, we
introduce two sets of QPM structures with different periodicities to
compensate for the phase mismatch between the FF and SH components, as well
as the mismatch between the TH component and the FF-SH complex. Numerical
simulations demonstrate that, when a Laguerre-Gaussian (LG) beam, carrying
the orbital angular momentum with winding number $l=1$, is injected into the
nonlinear crystal, the cascading effect not only facilitate THG but also
stabilizes vortex solitons.

This study provides new insights into the creation of vortex solitons in
pure $\chi ^{(2)}$ media and offers novel possibilities for the
vortex-soliton control and development of feasible applications, such as
those which may be promising for optical information processing. Below, we
first introduce the theoretical model in Section 2, followed by the
presentation and analysis of results of systematic simulations in Section 3
and estimation of necessary experimental parameters in Section 4. The paper
is concluded by Section 5.

\section{The model}

\begin{figure}[tbp]
{\includegraphics[trim=40 150 0 140,clip,width=3.8in]{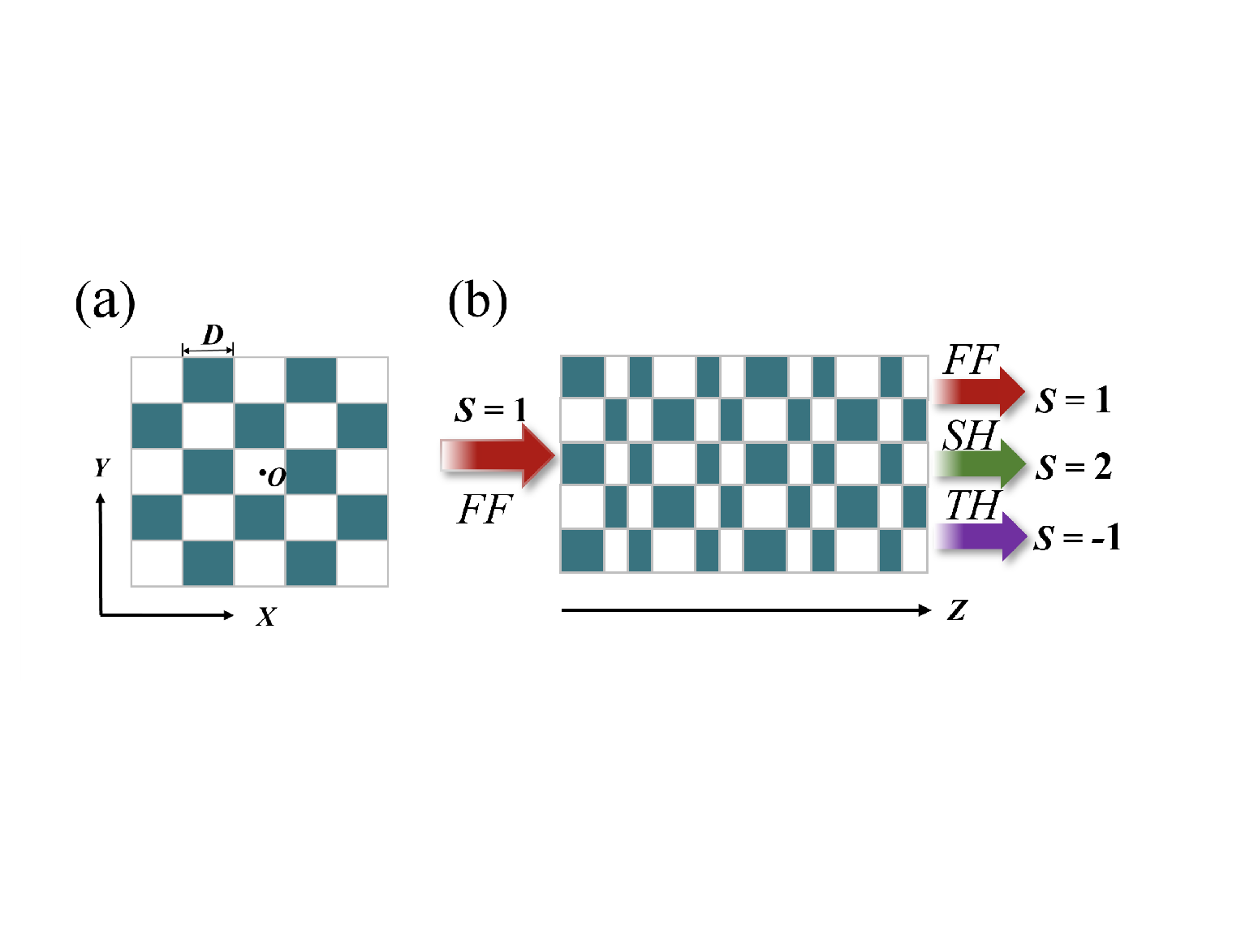}}
\caption{The schematic of the nonlinear optical lattice structure and
wave-field propagation. (a) The checkerboard-patterned structure in the
transverse $(X,Y)$ plane at $z=0$, with the distribution of the $\protect%
\chi ^{(2)}$ coefficient following Eq. (\protect\ref{sigma}). Dark green and white lattice cells correspond to regions with $\protect\sigma (x,y)=+1$ and $%
\protect\sigma (x,y)=-1$, respectively. (b) A schematic illustration of the
propagation in the nonlinear lattice along the $z$-direction. The red arrow
on the left represents the injection of a fundamental wave (FF) with winding
number (topological charge) $1$. The arrows at the right output end indicate
different harmonic components: FF (red), SH (green), and TH (purple), with
winding numbers $+1$, $+2$, and $-1$, respectively.}
\label{model}
\end{figure}

The paraxial propagation of the FF, SH, and TH waves, with the carrier
frequencies $\omega _{1,2,3}$ and slowly varying envelopes $A_{1,2,3}$, $\ $%
along the $Z$-direction in the crystals with the superlattice modulation is
governed by the coupled equations:
\begin{equation}
\begin{split}
i\partial _{Z}A_{1}=& -\frac{1}{2k_{1}}\nabla ^{2}A_{1}-\frac{%
2d(X,Y,Z)\omega _{1}}{cn_{1}} \\
& \left( A_{1}^{\ast }A_{2}e^{-i\Delta k_{a}Z}+A_{2}^{\ast }A_{3}e^{-i\Delta
k_{b}Z}\right) ,
\end{split}
\label{oh1}
\end{equation}%
\begin{equation}
\begin{split}
i\partial _{Z}A_{2}=& -\frac{1}{2k_{2}}\nabla ^{2}A_{2}-\frac{%
2d(X,Y,Z)\omega _{2}}{cn_{2}} \\
& \left( \frac{1}{2}A_{1}^{2}e^{-i\Delta k_{a}Z}+A_{1}^{\ast
}A_{3}e^{-i\Delta k_{b}Z}\right) ,
\end{split}
\label{oh2}
\end{equation}

\begin{equation}
i\partial _{Z}A_{3}=-\frac{1}{2k_{3}}\nabla ^{2}A_{3}-\frac{2d(X,Y,Z)\omega
_{3}}{cn_{3}}A_{1}A_{2}e^{i\Delta k_{b}Z},  \label{oh3}
\end{equation}%
where $c$ is the speed of light in vacuum, $n_{1,2,3}$ are values of the
refractive index corresponding to $\omega _{1,2,3}$,
the carrier frequencies are $\omega_j$=$j \omega_1$, $j=1,2,3$, the respective
wavenumbers are $k_{j}={n_{j}\omega _{j}}/{c}$, and the phase mismatches are
\begin{equation}
\Delta k_{a}=2k_{1}-k_{2},~\Delta k_{b}=k_{1}+k_{2}-k_{3}.  \label{ab}
\end{equation}%
The spatially distributed $\chi ^{(2)}$ coefficient is $d(X,Y,Z)=\sigma
(X,Y)d(Z) $, where $\sigma (X,Y)$ is shaped as a checkerboard pattern in the
transverse $(X,Y)$ plane:
\begin{equation}
\sigma (X,Y)=-\text{sgn}\left\{ \cos \left( \frac{\pi X}{D}\right) \cos
\left( \frac{\pi Y}{D}\right) \right\} ,  \label{sigma}
\end{equation}%
Each square unit, determined by Eq. (\ref{sigma}), has a size of $D\times D$%
, as shown in Fig. \ref{model}(a) \cite{jol2004}. The modulation function of
the $\chi ^{(2)}$ nonlinear coefficient along the propagation direction is
defined by
\begin{equation}
d(Z)=d_{0}\{\mathrm{sgn}[\cos (2\pi Z/\Lambda _{a})]+\mathrm{sgn}[\cos (2\pi
Z/\Lambda _{b})]\},  \label{d}
\end{equation}%
which is the superposition of two different poling periodicities, $\Lambda
_{a}$ and $\Lambda _{b}$, with amplitude $d_{0}$. Expression (\ref{d}) can
be represented by its Fourier expansion \cite{aoe2018,afp2021}:
\begin{align}
d(Z)=&d_{0}\Bigg[ \sum_{m=-\infty }^{+\infty }\frac{2}{m \pi}\sin \left(\frac{m \pi}{2}\right)\exp \left( i\frac{2\pi m}{\Lambda _{a}}Z\right) \notag\\
&+\sum_{l=-\infty }^{+\infty }\frac{2}{l \pi}\sin \left(\frac{l \pi}{2}\right)\exp \left( i \frac{2\pi l}{\Lambda _{b}}Z\right) \Bigg], \label{d(Z)}
\end{align}
%
%

Equations (\ref{oh1})-(\ref{oh3}) can be essentially simplified in the
near-resonance case, when terms with $m=\pm 1$ and $l=\pm 1$ in Eq. (\ref%
{d(Z)}) oscillate with the spatial frequencies close to $\Delta k_{a,b}$. In
this case, we neglect rapidly oscillating terms and introduce notation (cf.
Refs. \cite{cjosab2013,foe2021,ypra2020,gjosab2000,jpla2021})

\begin{equation}
I_{0}=\left( \frac{n_{1}}{\omega _{1}}+\frac{n_{2}}{\omega _{2}}+\frac{n_{3}%
}{\omega _{3}}\right) |A_{0}|^{2},  \label{I0}
\end{equation}%
where \(A_0\) is a characteristic amplitude of the electromagnetic field,
\begin{equation}
\begin{split}
& u_{j}=\left( \frac{\omega _{1}}{n_{1}I_{0}}\right) ^{-\frac{1}{2}%
}A_{j}e^{i\left( \Delta k_{a}-\frac{2\pi }{\Lambda _{a}}\right) Z}\quad
\text{for }j=1,2, \\
& u_{3}=\left( \frac{\omega _{1}}{n_{1}I_{0}}\right) ^{-\frac{1}{2}}A_{3}e^{i%
\left[ (2\Delta k_{a}-\Delta k_{b})-2\pi \left( \frac{2}{\Lambda _{a}}-\frac{%
1}{\Lambda _{b}}\right) \right] Z},
\end{split}
\label{uj}
\end{equation}%
\begin{equation}
\begin{split}
& z_{a}^{-1}=\frac{2d_{0}}{\pi c}\left( \frac{\omega _{1}^{2}\omega _{2}}{%
n_{1}^{2}n_{2}}I_{0}\right) ^{\frac{1}{2}}, \\
& z_{b}^{-1}=\frac{2d_{0}}{\pi c}\left( \frac{\omega _{1}\omega _{2}\omega
_{3}}{n_{1}n_{2}n_{3}}I_{0}\right) ^{\frac{1}{2}},
\end{split}
\label{zazb}
\end{equation}%
\begin{equation}
\kappa =\frac{z_{b}}{z_{a}}=\sqrt{\frac{n_{3}}{3n_{1}}},  \label{kappa}
\end{equation}

\begin{equation}
\Omega _{a}=z_{a}\left( \Delta k_{a}-\frac{2\pi }{\Lambda _{a}}\right)
,\quad \Omega _{b}=z_{a}\left( \Delta k_{b}-\frac{2\pi }{\Lambda _{b}}%
\right) ,  \label{omegaa}
\end{equation}

\begin{equation}
\Omega _{3}=2\Omega _{a}-\Omega _{b},  \label{omega3}
\end{equation}
and the simplified equations are cast in the scaled form:
\begin{equation}
i\partial _{z}u_{1}=-\frac{1}{2}\nabla ^{2}u_{1}-\Omega _{a}u_{1}-\sigma
(x,y)\left( 2u_{1}^{\ast }u_{2}+2\kappa u_{2}^{\ast }u_{3}\right)
\label{gyh1}
\end{equation}%
\begin{equation}
i\partial _{z}u_{2}=-\frac{1}{4}\nabla ^{2}u_{2}-\Omega _{a}u_{2}-\sigma
(x,y)\left(u_{1}^{2}+2\kappa u_{1}^{\ast }u_{3}\right)  \label{gyh2}
\end{equation}%
\begin{equation}
i\partial _{z}u_{3}=-\frac{1}{6}\nabla ^{2}u_{3}-\Omega _{3}u_{3}-2\sigma
(x,y)\kappa u_{1}u_{2}  \label{gyh3}
\end{equation}

According to the Manley-Rowe relations \cite{gjosab2013}, the system
possesses two dynamical invariants, \textit{viz}., the total power
\begin{equation}
P=\int \int \left( |u_{1}|^{2}+2|u_{2}|^{2}+3|u_{3}|^{2}\right) dxdy
\label{power}
\end{equation}%
and Hamiltonian.
\begin{equation}
H=\int \int \left( \mathcal{H}_{P}+\mathcal{H}_{\Omega }+\mathcal{H}%
_{d}\right) \,dx\,dy,  \label{hmdl}
\end{equation}%
where the individual energy-density terms are defined as follows:
\begin{equation}
\mathcal{H}_{P}=\frac{1}{2}|\nabla u_{1}|^{2}+\frac{1}{4}|\nabla u_{2}|^{2}+%
\frac{1}{6}|\nabla u_{3}|^{2},  \label{hmdlp}
\end{equation}%
\begin{equation}
\mathcal{H}_{\Omega }=-\Omega _{a}(|u_{1}|^{2}+|u_{2}|^{2})-\Omega
_{3}|u_{3}|^{2},  \label{hmdlomega}
\end{equation}%
\begin{equation}
\mathcal{H}_{d}=-\sigma (x,y)\left[ g_{a}(u_{1}^{2}u_{2}^{\ast }+\mathrm{c.c}%
)+2g_{b}(u_{1}u_{2}u_{3}^{\ast }+\mathrm{c.c)}\right] ,  \label{hmdld}
\end{equation}%
where $\mathrm{c.c.}$ stands for the complex conjugate. The control
parameters of the system are $P$, $D$, $\Omega _{a}$,$\Omega _{b}$ and $\Omega_{3}$.
The vortex solitons are obtained below by means of the imaginary-time propagation method \cite{fylpra2008} \cite{sodpre2006}.

\section{Results and discussion}

Bright vortex-soliton solution of Eqs. (\ref{gyh1})--(\ref{gyh3}) are looked
for as
\begin{equation}
u_{p}(x,y,z)=\phi _{p}(x,y)\exp (i\beta _{p}z),\quad p=1,2,3,  \label{gzj}
\end{equation}%
where, $\phi _{1,2,3}$ and $\beta _{1,2,3}$ represent the steady-state
profiles and propagation constants of the respective components, the
propagation constants being subject to the obvious relation, $\beta
_{3}=\beta _{1}+\beta _{2}$. Further, phases of the complex components, $%
\varphi _{1,2,3}(x,y)=\text{Arg}\{\phi _{1,2,3}(x,y)\}$, are related by the
phase-matching conditions $%
\text{Arg}\{\phi _{3}(x,y)\}=\text{Arg}\{\phi _{1}(x,y)\}+\text{Arg}\{\phi
_{2}(x,y)\}$.

\begin{equation}
\varphi _{2}(x,y)=2\varphi _{1}(x,y)-\varphi _{d}(x,y),  \label{phase2}
\end{equation}%
\begin{equation}
\varphi _{3}(x,y)=\varphi _{1}(x,y)+\varphi _{2}(x,y)-\varphi _{d}(x,y),  \label{phase3}
\end{equation}%
where $\varphi _{d}(x,y)=0$ when $\sigma (x,y)=1$, and $\varphi
_{d}(x,y)=\pi $ when $\sigma (x,y)=-1$, see Eq. (\ref{sigma}).

\begin{figure}[tbp]
{\includegraphics[trim=110 75 0 75,clip,width=4.2in]{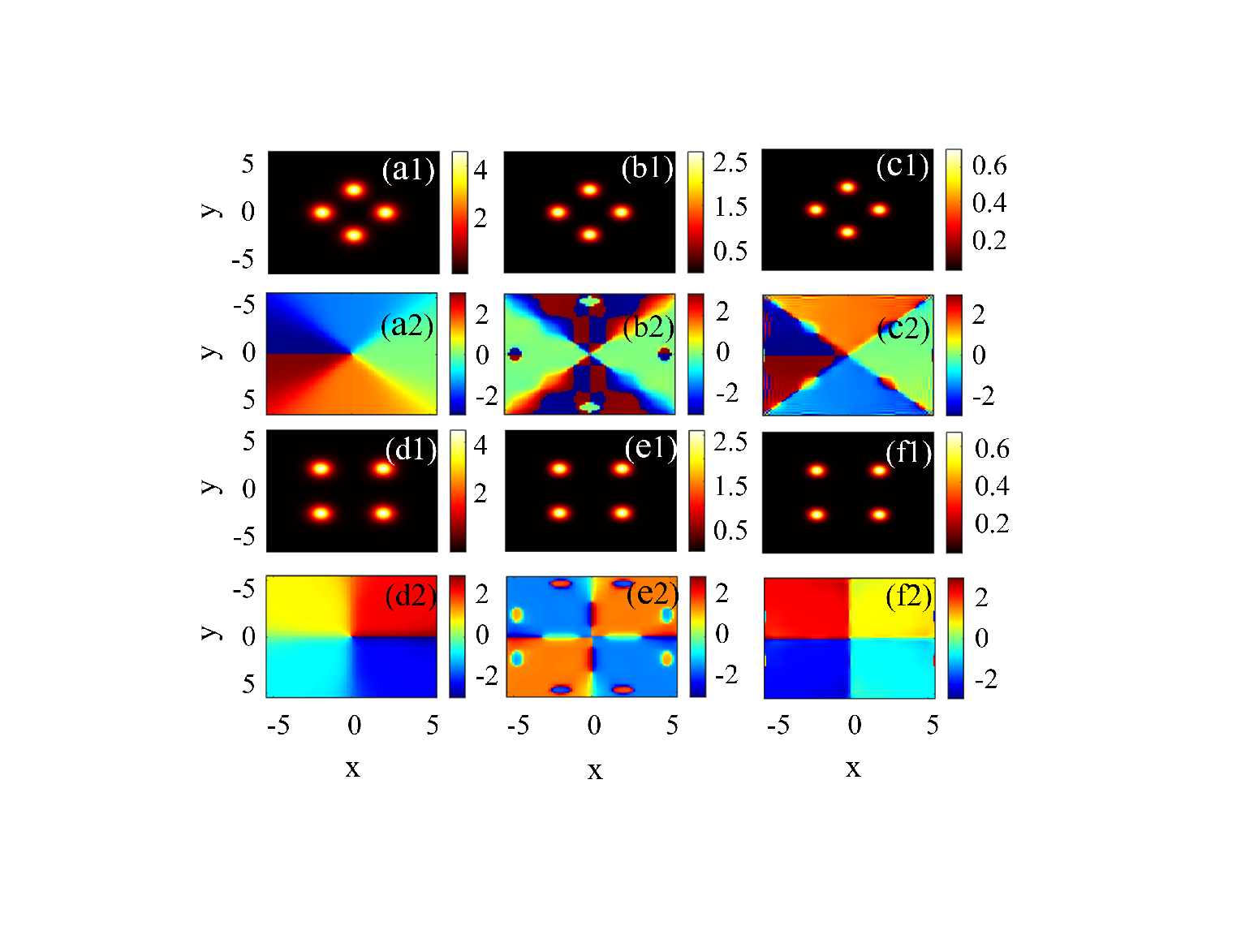}}
\caption{Examples of stable rhombic and square-shaped vortex solitons, each
built of four intensity peaks. Panels (a1)-(c1) and (a2)-(c2), and (d1)-(f1)
and (d2)-(f2) display the intensity and phase distributions of the FF, SH,
and TH components invortex solitons, respectively. The phase-matching
conditions (\protect\ref{phase2}) and (\protect\ref{phase3}) for the three
components of the rhombic-shaped soliton at the peak locations amount to $%
\protect\varphi _{2}(x,y)=2\protect\varphi _{1}(x,y)$ and $\protect\varphi %
_{3}(x,y)=\protect\varphi _{1}(x,y)+\protect\varphi _{2}(x,y)$, while for
the square-shaped soliton they are $\protect\varphi _{2}(x,y)=2\protect%
\varphi _{1}(x,y)-\protect\pi $ and $\protect\varphi _{3}(x,y)=\protect%
\varphi _{1}(x,y)+\protect\varphi _{2}(x,y)$. The central regions of the
rhombic and square patterns correspond to the white square in the
checkerboard structure shown in Figs. \protect\ref{model}. The parameters
used here are $(P,D,\Omega _{a},\Omega _{b},\Omega _{3})=(50,4,0,0,0)$.}
\label{qdfb}
\end{figure}

Numerical results demonstrate the existence of two distinct families of
vortex solitons with different shapes, each composed of four local intensity
peaks: rhombic ones, as shown in Figs. \ref{qdfb}(a-c), and square-shaped
vortices, which are shown in Figs. \ref{qdfb}(d-f). These two families
correspond, respectively, to the phase-matching conditions (\ref{phase2})
and (\ref{phase3}) with $\varphi _{d}(x,y)=0$ and $\varphi _{d}(x,y)=\pi $, at positions of the
intensity peaks. The rhombic and square configurations exhibit, severally,
more compact and more sparse spatial distributions. The vortex-soliton modes
found here do not feature visible side peaks outside the rhombic or square
patterns, which makes them markedly different from multi-peak vortex
solitons produced by three-wave systems in Refs. \cite{gjosab2000,jpla2021}.
This sharp structure of the rhombic and square-shaped vortices  may be
beneficial for potential applications.

The phase distributions of the rhombic and square solitons shown in Fig. \ref%
{qdfb} reveal their vortex topologies, which are defined phase circulation
along closed paths connecting the intensity peaks, even if the angular
momentum is not conserved. This is the standard definition for
characterizing multi-peak vortices in systems with matter-wave systems with
optical-lattice potentials \cite{bel2003,jol2003}. The consideration of the
phase profiles demonstrates that the FF components of both rhombic and
square solitons (Figs. \ref{qdfb} (a2, d2)) exhibit vortex structures with
the winding number (topological charge) $s=1$. The SH components (Figs. \ref%
{qdfb} (b2, e2)) features its topological charge, $s=2$. As the displayed phase range is restricted to $%
|\varphi _{2}|<2\pi $, local phase values exceeding $2\pi $ are
reconstructed by subtracting $2\pi $, resulting in the quadrupole-like phase
distribution for the SH, as shown in Eq. (\ref{Quadrupole}). For the TH
components, whose phase values are restricted to the same range, we note
that the $D_{4}$ symmetry group of the underlying photonic lattice does not
support vortex states with the topological charge greater than one \cite%
{aprl2005}. Therefore, the TH components (see Figs. \ref{qdfb} (c2,f2))
exhibit vortex structures with topological charge $s=-1$, characterized
by the phase circulation opposite to that of the FF field. We refer to it as
an anti-vortex structure, as illustrated by the following schematic relations:

\begin{eqnarray}
\text{FF} &\otimes& 2\Longrightarrow \text{SH}\notag \\
(\text{FF Vortex}) &\otimes& 2\Longrightarrow \text{SH\ Quadrupole,}  \label{Quadrupole}
\end{eqnarray}%
\begin{eqnarray}
\text{FF} \oplus \text{SH}\Longrightarrow \text{TH}&\notag \\
(\text{FF Vortex})\oplus (\text{SH\ Quadrupole}) &\Longrightarrow \text{Anti-vortex.} \label{Anti-vortex}
\end{eqnarray}%
These results indicate that the vortex structure in the FF component, with $%
s=1$, enables the effective generation of the anti-vortex structure, with $%
s=-1$, in TH via the nonlinear coupling mediated by the quadrupole mode of
the SH component ($s=2$) in the nonlinear photonic crystal. This mechanism
suggests a novel pathway for the construction and control of higher-order
vortex states in nonlinear optics.

\begin{figure}[tbp]
{\includegraphics[trim=30 105 70 135,clip,width=3.4in]{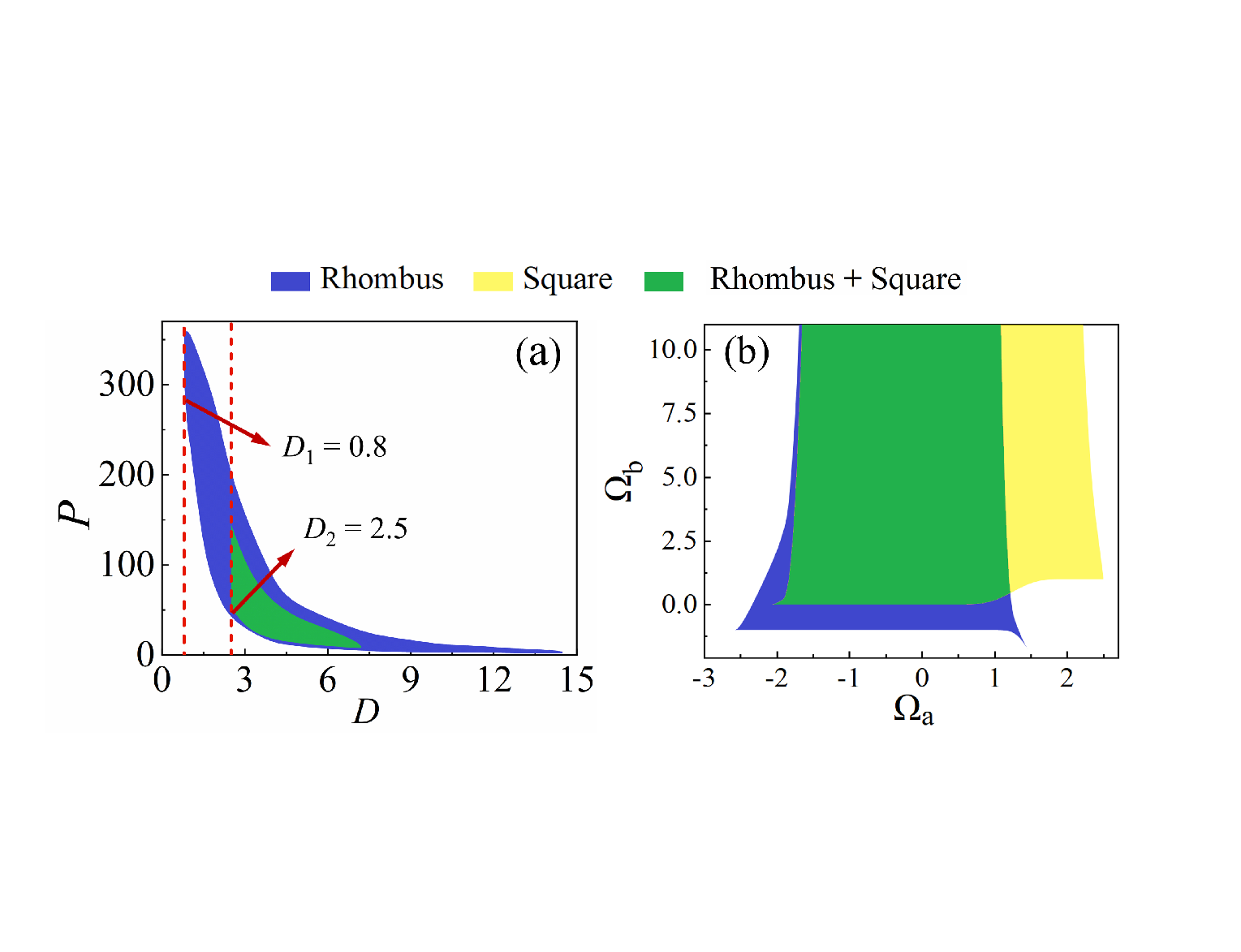}}
\caption{The rhombic and square-shaped vortex solitons are stable, respectively, in the blue and yellow regions of the $(P,D)$ parameter plane for $(\Omega_{a}, \Omega_{b}, \Omega_{3}) = (0,0,0)$ (panel a), and in the $(\Omega_{a},\Omega_{b})$ plane for $(P,D) = (50,4)$, with $\Omega_{3} = 2\Omega_{a} - \Omega_{b}$ (panel b). The two soliton species coexist as stable solutions in the green regions in both parameter planes. In panel (a), the vertical red dotted lines mark the existence boundaries for the two types of vortex solitons.}
	\label{qyt}
\end{figure}


We verified the stability of the vortex solitons by simulations of their perturbed evolution
in the framework of Eqs. (\ref{gyh1})--(\ref{gyh3}), over the distance $z=1000$ (it corresponds
to  $\simeq 100$ diffraction lengths of the characteristic soliton states displayed in Fig. \ref{qdfb},
cf. Refs. \cite{kprl2017,jprl2003}). The results of the systematic simulations are  summarized in Fig. \ref{qyt}(a)
in the form of the stability regions of the rhombic and square-shaped vortex solitons in the parameter plane $(P,D)$,
with the other parameters fixed as $(\Omega _{a},\Omega _{b},\Omega _{3})=(0,0,0)$. In this figure, the blue region represents
the stability domain of the rhombic vortex solitons, while the green region indicates the coexistence of both rhombic and square-shaped vortex solitons.
Thus, the rhombic ones are stable in an essentially broader
area. In particular, the rhombic solitons maintain their stability for the
total power of up to $P_{\text{max}}\approx 350$, whereas the square-shaped solitons
remain stable only up to $P_{\text{max}}\approx 140$. Outside of the stability regions, unstable solutions have not been found.
Furthermore, the stability regions of the rhombic and square-shaped vortex solitons are plotted in the $(\Omega_a, \Omega_b)$ parameter plane in Fig. \ref{qyt}(b),
with other parameters fixed as $(P,D) = (50,4)$ and $\Omega_{3} = 2\Omega_{a} - \Omega_{b}$.
Unstable soliton solutions exist outside of the left and right boundaries of the stability regions for both types of the
solitons. When $\Omega_b$ is smaller than its value at the lower boundary of the stability domains for both types of the solitons,
no vortex soliton solutions exist; on the other other hand, the stability domain extends towards $\Omega_b\rightarrow\infty$, and the
energy carried by the third-harmonic component decreases for both types of the solitons  with the increase of $\Omega_{b}$.

The minimum spatial lattice period $D_{\text{min}}$ in Eq. (\ref{sigma}),
which is required to support the rhombic and square-shaped solitons, is
indicated by the vertical red dashed lines in Fig. \ref{qyt}. It is $%
D_{1}=0.8$ and $D_{2}=2.5$ for the rhombic and square-shaped solitons,
respectively. In the case of $D<D_{\text{min}}$, the rapid alternation of
the sign of the $\chi ^{(2)}$ terms in Eqs. (\ref{gyh1})--(\ref{gyh3}) leads
to an effective cancellation of the quadratic nonlinearity, resulting in the
nonexistence of vortex solitons. At $D>D_{\text{min}}$, those rhombic or
square-shaped solitons which are unstable spontaneously decay into one or
two localized soliton peaks, respectively.
Thus, we conclude that the compact structure of the rhombic solitons makes
it easier to maintain their stability, in comparison to the loosely coupled
square-shaped solitons.

\begin{figure}[h]
{\includegraphics[trim=130 82 70 110,clip,width=4in]{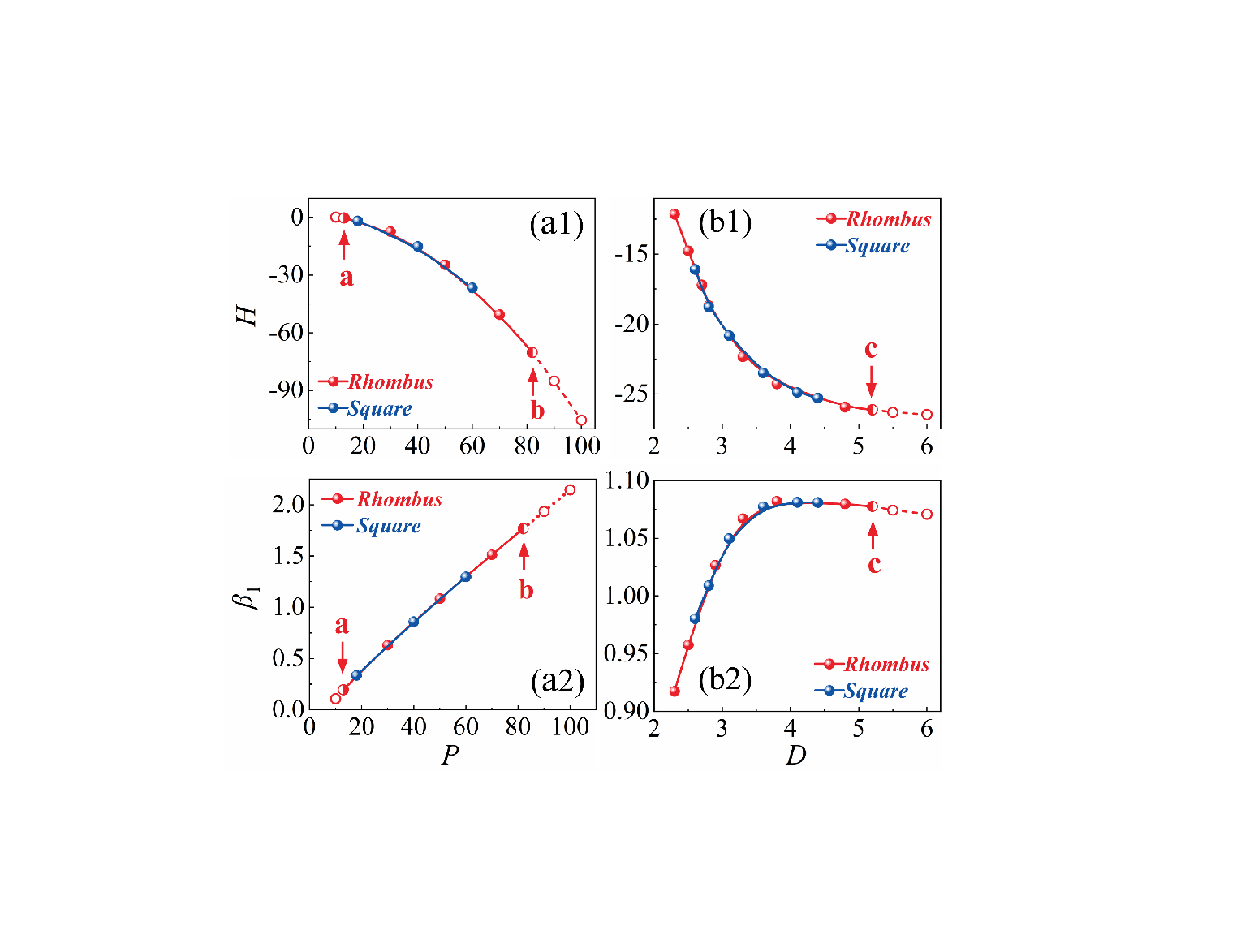}}
\caption{Dependences of Hamiltonian $H$ and propagation constant $\protect%
\beta $ of the two types of\ the vortex solitons on the control parameters,
\textit{viz}., total power $P$ and lattice period $D$, respectively. Solid lines represent
stable rhombic and square-shaped solitons, while the dashed lines correspond
to unstable rhombic-shaped modes. Red and blue spheres are data points
denoting stable rhombic-shaped and square-shaped solitons, respectively. Red
circles indicate unstable rhombic-shaped solitons. Point $a$ in panels (a1)
and (a2) marks the stability boundary between the stable and unstable
rhombic-shaped solitons. Points $b$ in (a1) and (a2), and point $c$ in (b1)
and (b2), represent the boundary between stable and unstable rhombic
solitons. The parameters are $(D,\Omega _{a},\Omega _{b},\Omega
_{3})=(4,0,0,0)$ for (a1) and (a2), and $(P,\Omega _{a},\Omega _{b},\Omega
_{3})=(50,0,0,0)$ for (b1) and (b2).}
\label{pdxzt}
\end{figure}

For both types of the vortex solitons considered here, we conducted a
detailed analysis of the dependence of their characteristics, including
Hamiltonian $H$ and propagation constant $\beta $ (see Eqs. (\ref{hmdl}) and
(\ref{gzj}), on the control parameters, i.e., total power $P$ and lattice
period $D$. The results are presented in Fig. \ref{pdxzt}. once again for
other parameters fixed as $(\Omega _{a},\Omega _{b},\Omega _{3})=(0,0,0)$.
At a fixed lattice period of $D = 4$, the rhombic-shaped vortex solitons exist for total powers up to $P \leq 100$,
whereas the square-shaped vortex solitons are limited to $P \leq 60$ (Figs. \ref{pdxzt}(a1),(a2)). Conversely, at a
fixed total power of $P = 50$, the rhombic-shaped vortex solitons exist for lattice periods up to $D \leq 6$, while
the square-shaped ones are confined to $D \leq 4.4$ (Figs. \ref{pdxzt}(b1),(b2)).
Notably, under these conditions, all square-shape solitons are found to be stable. The results summarized in Figs. \ref{pdxzt}%
(a1,b1) reveal that the rhombic solitons exhibit a broader tunability in the
parameter space and possess significantly lower values of the Hamiltonian
values than the square-shaped solitons. The latter feature explains the
broader stability region of the rhombic solitons observed in Fig. \ref{qyt},
as they are more favorable energetically. Note that the $\beta (P)$ curves,
plotted in Figs. \ref{pdxzt}(a2), demonstrate that both soliton families
satisfy the Vakhitov-Kolokolov (VK) stability criterion, $d\beta /dP>0$ \cite%
{nqe1973}, which is a well-known necessary condition for the stability of
soliton solutions in nonlinear systems.

\begin{figure}[tbp]
{\includegraphics[trim=122 80 0 80,clip,width=4.4in]{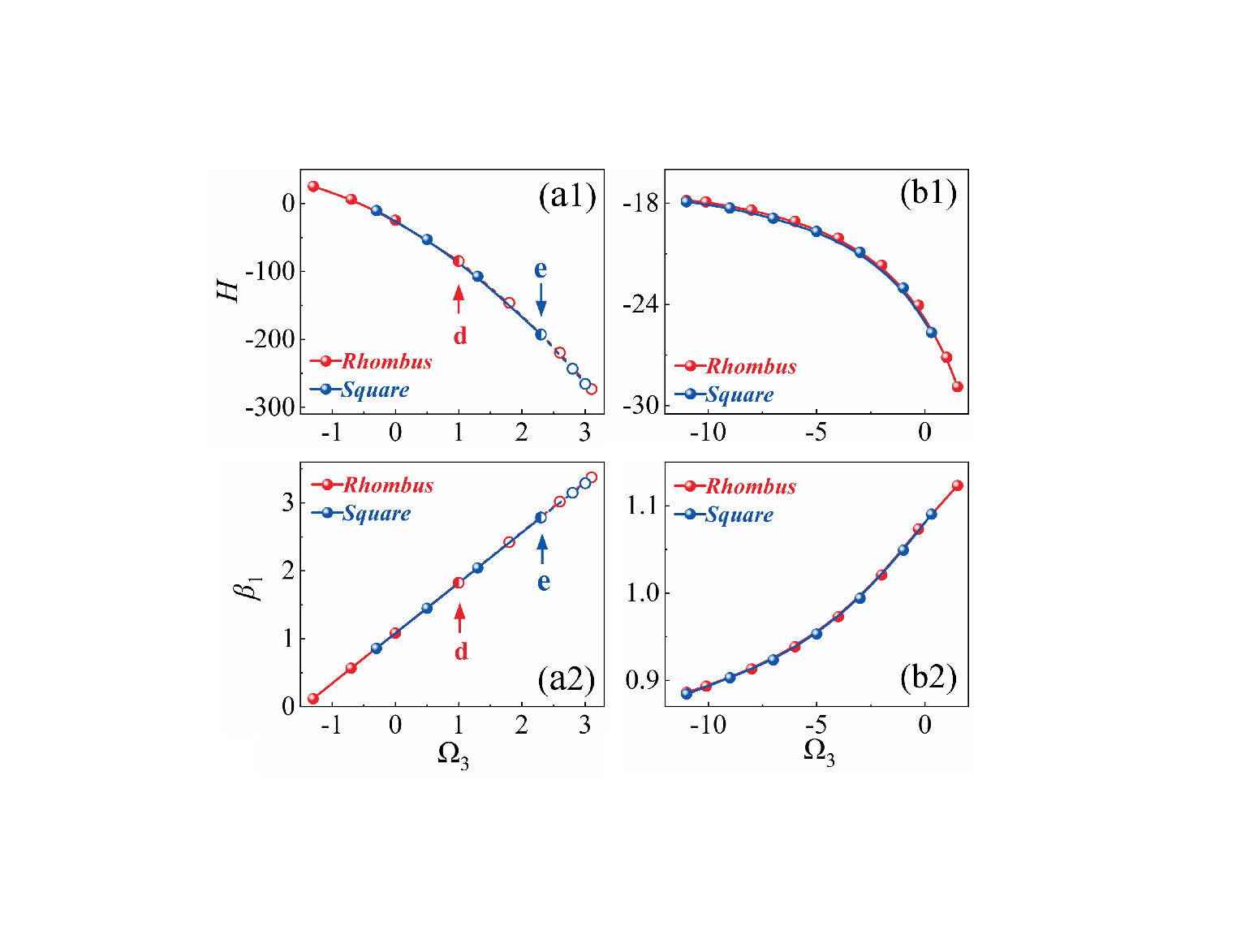}}
\caption{Hamiltonian $H$ and propagation constant $\protect\beta $ of the
two rhombic and square-shaped vortex solitons vs. detuning parameters $%
\Omega _{a}$ and $\Omega _{3}$. Solid and dashed lines indicate,
respectively, stable and unstable subfamilies of the solitons. regions of
vortex solitons, Red and blue spheres denote stable rhombic and
square-shaoed solitons, respectively, whereas red and blue circles indicate
unstable ones. Points $d$ and $e$ in (a1) and (a2) mark boundaries between
stable and unstable rhombic and square-sahped solitons, respectively.
Parameters for (a1) and (a2) are $\Omega _{a}=\Omega _{b}=\Omega _{3}$, $%
(P,D)=(50,4)$, whereas for (b1) and (b2) they are $\Omega _{3}=-\Omega _{b}$%
, $(P,D,\Omega _{a})=(50,4,0)$.}
\label{omegaxzt}
\end{figure}

Due to the presence of two distinct periodic modulation structures in the
system, two different detuning parameters, $\Omega _{a}$ and $\Omega _{3}$,
are introduced in Eqs. (\ref{omegaa}) and (\ref{omega3}). To demonstrate
effects of detuning on the solitons, we plots dependences of $H$ and $\beta $
on the $\Omega _{a}$ and $\Omega _{3}$ in Fig. \ref{omegaxzt}. In the case
of $\Omega _{a}=\Omega _{b}=\Omega _{3}$, and $(P,D)=(50,4)$, the numerical
results reveal that the rhombic solitons produce lower values of the
Hamiltonian values, enabling their stable existence in a broader parameter
range. Note also that both soliton types satisfy the VK criterion. Further,
it is worthy to note that, at $\Omega _{a}=0$, both solitons are stable for
these parameters, yet the rhombic one demonstrates a larger existence range.
The results indicate that the detuning significantly affects the soliton
stability. Due to their stronger structure, the rhombic solitons exhibit
enhanced stability under various detuning conditions. These findings provide
relevant insights for the control and possible application of vortex
solitons in nonlinear systems.

\begin{figure}[tbp]
{\includegraphics[trim=90 55 0 70,clip,width=4.3in]{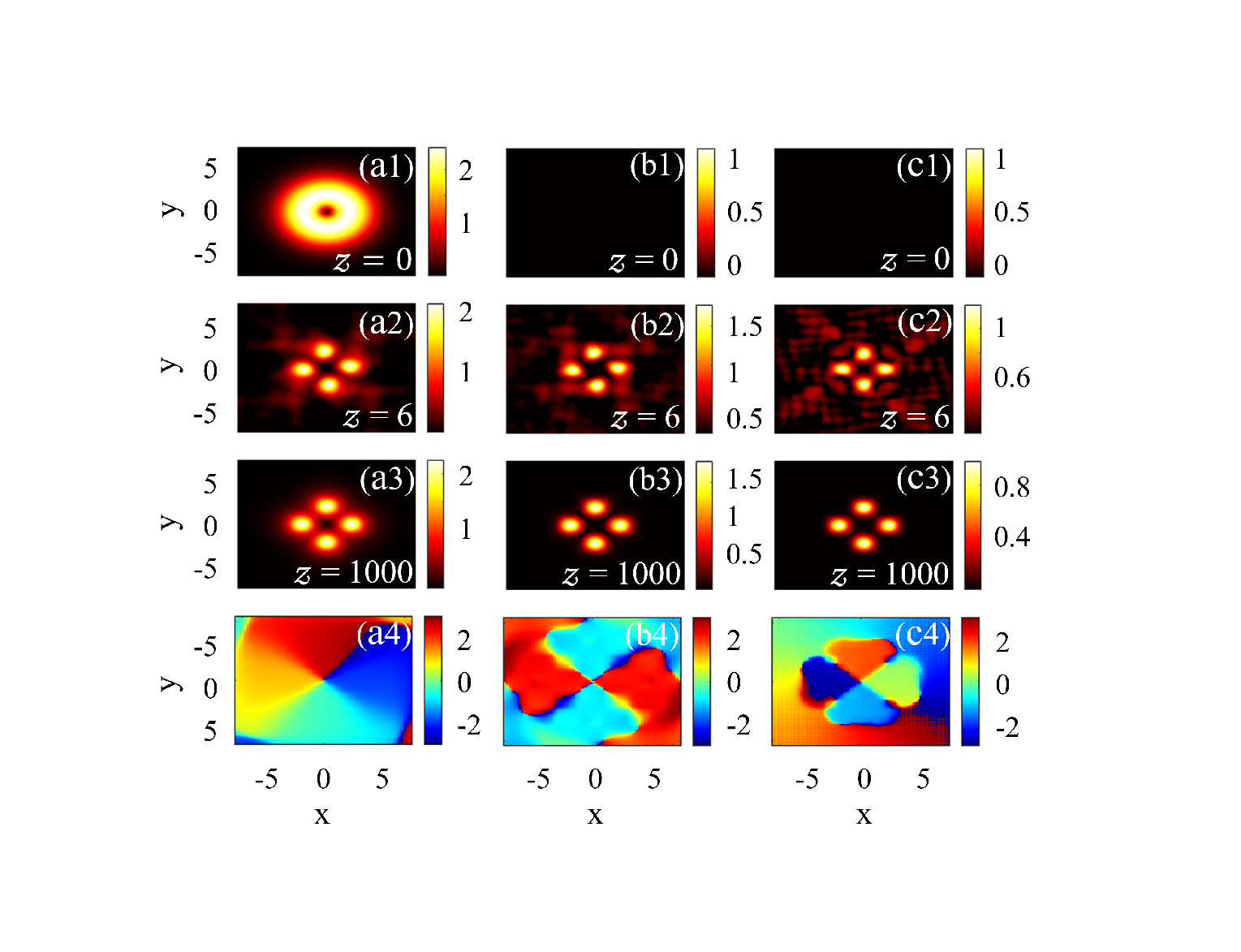}}
\caption{A typical example of the simulated creation of a rhombic vortex
soliton in a lithium niobate crystal. At the initial position $z=0$, the FF
component is launched as an LG vortex beam with power $P=100$ and
topological charge $1$. Panels (a1)-(a3), (b1)-(b3), and (c1)-(c3) display
the intensity distributions of the FF, SH, and TH components, respectively,
at values of the propagation distance indicated in the panels, while the
phase distributions in the three components at $z=1000$ is displayed in
(a4)-(c4). The parameters are $D=3$ and $(\Omega _{a},\Omega _{b},\Omega
_{3})=(0,0,0)$.}
\label{rhombic}
\end{figure}

\begin{figure}[tbp]
{\includegraphics[trim=85 50 0 33,clip,width=4.2in]{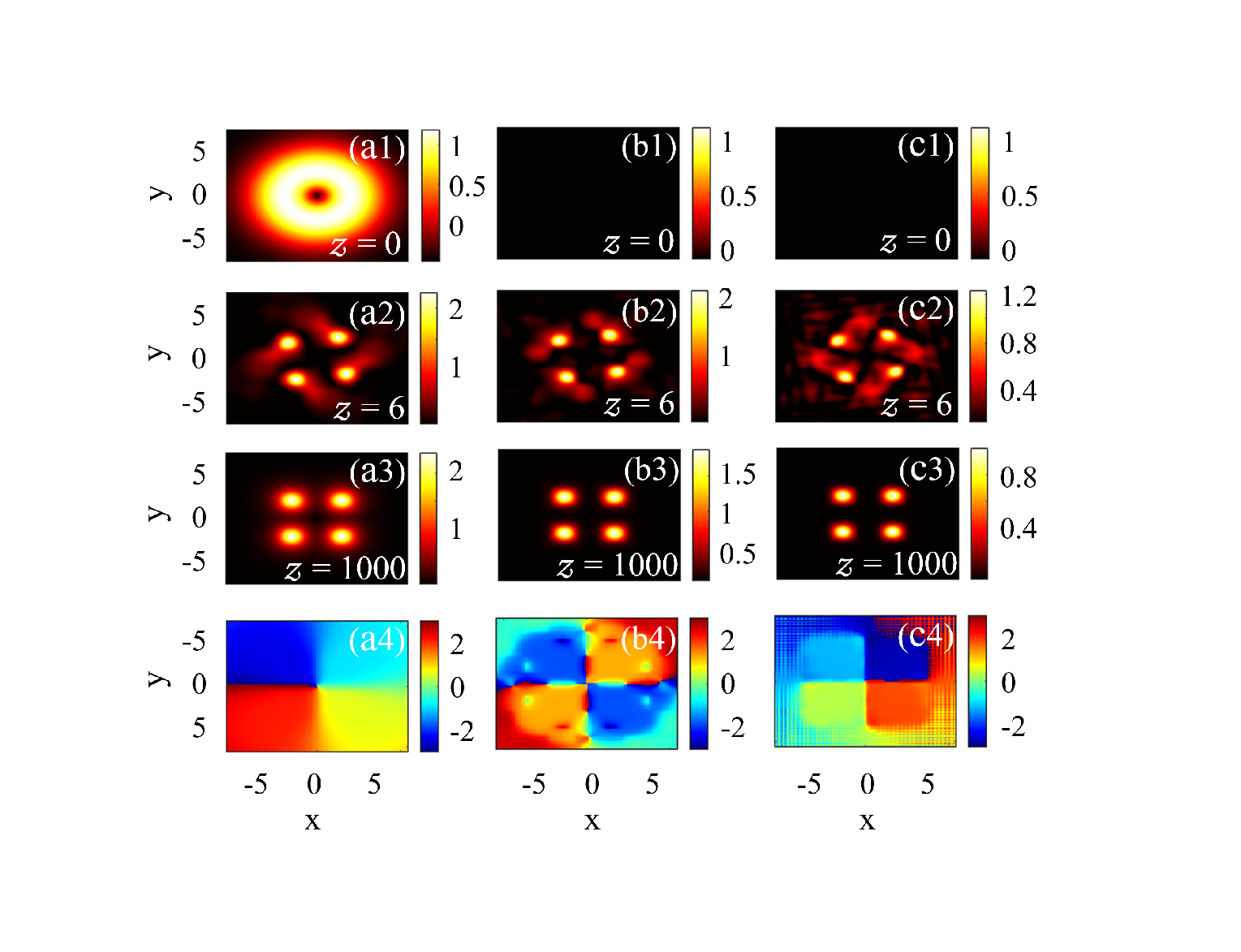}}
\caption{A typical example of the simulated creation of a square-shaped
vortex soliton in the lithium niobate crystal. At the initial position $z=0$%
, the FF component is launched as an LG vortex beam with power $P=100$ and
topological charge $1$. Panels (a1)-(a3), (b1)-(b3), and (c1)-(c3) display
intensity distributions in the FF, SH, and TH components, respectively, at
values of $z$ indicated in the panel,  Panels (a4)-(c4) display the phase
distributions in the components at $z=1000$. The parameters are $D=3$ and $%
(\Omega _{a},\Omega _{b},\Omega _{3})=(0,0,0)$.}
\label{square}
\end{figure}

To realize the excitation of different types of the  three-component vortex
solitons, we employ LG beams with varying beam widths as the input fields
for the FF component, while the initial field amplitudes of the SH and TH
components are zero. Simulations reveal that broader input beams tend to
generate more extended spatial structures, leading to the formation of
vortex solitons with square-shaped geometries. Conversely, narrower beams
result in more localized initial intensity distributions, favoring the
excitation of compact rhombic vortex solitons.
As shown in Figs. \ref{rhombic} and \ref{square}, both the rhombic and square-shaped vortex
solitons undergo structural evolution and establish a well-defined shape at
the propagation distance $z=6$. They remain stable, at least, up to $z=1000$.
Moreover, stable propagation is maintained even at $z = 1300$, which corresponds to a physical distance of $1$ m. indicating
the robust long-distance transmission of the solitons. We have also found that
the propagation distance required to complete the soliton formation significantly decreases with the increase of the input power.

These results indicate that the transverse width of the input FF beam plays
a crucial role in determining the shape of the resulting soliton. By
appropriately tuning this parameter, one can selectively excite and control
either rhombic or square-shaped composite vortex solitons. To the best of
our knowledge, this study reports for the first time stable propagation of
square-shaped vortex solitons which combines the FF, SH, and TH components.

\section{Estimation of experimental parameters}

To assess the experimental feasibility of the proposed vortex solitons, we
conducted a systematic quantitative analysis based on the material
parameters of lithium niobate (LiNbO$_{3}$), taking the size of the $\chi
^{(2)}$ coefficient in Eq. (\ref{d}) as $d_{0}=27\,\mathrm{pm/V}$ \cite{http}%
. The wavelengths of the FF, SH, and TH components were selected as $\lambda
_{1}=1422\,\mathrm{nm}$, $\lambda _{2}=711\,\mathrm{nm}$, and $\lambda
_{3}=474\,\mathrm{nm}$, respectively. The electric-field amplitude was set
to $A_{0}=200\,\mathrm{kV/cm}$. To account for the dispersion in the
material, we adopted the wavelength-dependent refractive indices of LiNbO$%
_{3}$: $n_{1}\approx 2.2174$ (at 1422 nm), $n_{2}\approx 2.2709$ (at 711
nm), and $n_{3}\approx 2.3601$ (at 474 nm) \cite{djosab1997}. Under these
conditions, the conversion relations between the normalized variables and
physical units are summarized in Table \ref{tab:relation}. According to Eq. (%
\ref{zazb}), the characteristic propagation length is calculated as $%
z_{a}=0.0765\,\mathrm{cm}$. Based on the parameters in Table \ref%
{tab:relation}, the actual total power of the soliton in Fig. \ref{qdfb} is
estimated to be $\approx 3.8\,\mathrm{kW}$. The normalized propagation
distance $z=1000$ corresponds to length $76.5\,\mathrm{cm}$ in physical
units, which as mentioned above, $\simeq 100$ diffraction lengths. The
third-order nonlinear susceptibility of lithium niobate is $\chi
^{(3)}=36.6\times 10^{-22}\,\mathrm{m^{2}/V^{2}}$ \cite{ispinjose}. In Fig. %
\ref{qdfb}, the peak intensities of the FF, SH, and TH components are $%
\approx 0.42\,\mathrm{GW/cm^{2}}$, $0.52\,\mathrm{GW/cm^{2}}$, and $0.21\,%
\mathrm{GW/cm^{2}}$, respectively, all of which are significantly below the
threshold for triggering the Kerr nonlinearity ($1\simeq \,\mathrm{GW/cm^{2}}
$). Therefore, the Kerr nonlinearity may be neglected in the present
setting. It is worth to note that the spatial scale of the structures listed
in Table \ref{tab:relation} falls within fabrication capabilities of current
QPM techniques. These results not only uphold the experimental feasibility
of the proposed scheme but also offer theoretical guidance for the
optimization of parameters of possible applications.


\begin{table}[h]
\caption{The relationship between the scaled and physical units, assuming
refractive indices $n_{1}\approx 2.2174$, $n_{2}\approx 2.2709$, and $%
n_{3}\approx 2.3601$}\centering
\setlength{\arrayrulewidth}{0.8pt} 
\begin{tabular}{cc}
\hline\hline
$x = 1$ & 8.84 $\mu$m \\
$z = 1000$ & 76.5 cm \\
$P = 1$ & 77.44 W \\
$|u_j|^2 = 1, j = 1,2,3$ & \hspace{2em}99.2, 198.4, 297.6 MW/cm$^2$ \\
\hline\hline
\end{tabular}%
\label{tab:relation}
\end{table}

\section{Conclusion}

We have proposed a novel modulation scheme for the second-order nonlinear
susceptibility based on QPM\ (quasi-phase-matched) structures with two
different modulation periods. This design effectively compensates the phase
mismatch between the FF\ (fundamental-frequency), SH (second-harmonic), and
TH (third-harmonic) components, enabling the formation of two types of
four-lobed vortex soliton structures, with the rhombic and square shapes. We
have demonstrated, for the first time, the stable propagation of the
three-component vortex solitons. By means of the comprehensive numerical
analysis, we have systematically explored conditions for the formation of
stable composite vortex solitons in the three-dimensional QPM photonic
crystal, incorporating the TH generation through the cascading mechanism.
The corresponding stability regions in the parameter space were clearly
identified. The results reveal that the rhombic vortex solitons possess a
broader stability domain and lower values of the Hamiltonian, in comparison
to the square-shaped solitons, exhibiting superior structural robustness,
especially in the course of the long-distance propagation. Further numerical
analysis indicates that the propagation distance required for soliton
formation significantly decreases with the increase of the input power,
offering a mechanism for the rapid excitation of stable vortex solitons.
Moreover, quantitative estimates based on parameters of lithium niobate
confirm the experimental feasibility of the proposed soliton structures
under available technological conditions. In summary, this work not only
provides a novel theoretical framework for the excitation and control of
vortex solitons in nonlinear optics, but also offers insights for the
exploration of higher-order nonlinear phenomena and the design of novel
photonic-crystal devices.

As an extension of the present work, it may be relevant to consider parallel
copropagation of two or several vortex solitons of the same or different
types. In particular, such complexes may provide higher values of the
vorticity which, as mentioned above, cannot exceed $s=1$ for the single
vortex soliton.

\section*{Acknowledgments}

We appreciate valuable discussions with Guilong Li and Zibin Zhao. This work was supported by NNSFC (China)
through Grants No. 12274077, No. 12475014, the Natural Science Foundation of Guangdong province through
Grant No. 2025A1515011128, No. 2024A1515030131, No. 2023A1515110198,  No. 2023A1515010770, the Research Fund of Guangdong-Hong
Kong-Macao Joint Laboratory for Intelligent Micro-Nano Optoelectronic
Technology through grant No.2020B1212030010. The work of B.A.M. was supported, in part, by the
Israel Science Foundation through grant No. 1695/2022.


\begin{thebibliography}{99}
\bibitem{baip2022} B. A. Malomed, Multidimensional Solitons (AIP Publishing
LLC, 2022).

\bibitem{zcpb2012} Z. Guo and L. Ran, Rotational motions of optically
trapped microscopic particles by a vortex femtosecond laser, Chin. Phys. B
\textbf{21}, 104206 (2012).

\bibitem{wlpr2016} W. Yu, Z. Ji, D. Dong, X. Yang, Y. Xiao, Q. Gong, P. Xi,
and K. Shi, Super-resolution deep imaging with hollow Bessel beam STED
microscopy, Laser \& Photonics Reviews \textbf{10}, 147 (2016).

\bibitem{zsr2014} Z. Zhou, Y. Li, D. Ding, Y. Jiang, W. Zhang, S. Shi, B.
Shi, and G. Guo, Generation of light with controllable spatial patterns via
the sum frequency in quasi-phase matching crystals, Sci. Rep. \textbf{4},
5650 (2014).

\bibitem{jnp2012} J. Wang, J. Yang, I. M. Fazal, N. Ahmed, Y. Yan, H. Huang,
Y. Ren, Y. Yue, S. Dolinar, M. Tur, and A. E. Willner, Terabit free-space
data transmission employing orbital angular momentum multiplexing, Nature
Photonics \textbf{6}, 488 (2012).

\bibitem{apr2002} A. V. Buryak, P. D. Trapani, D. V. Skryabin, and S.
Trillo, Optical solitons due to quadratic nonlinearities: From basic physics
to futuristic applications, Phys. Rep. \textbf{370}, 63 (2002).

\bibitem{wprl1995} W. E. Torruellas, Z. Wang, D. J. Hagan, E. W.
VanStryland, G. I. Stegeman, L. Torner, and C. R. Menyuk, Observation of
Two-Dimensional Spatial Solitary Waves in a Quadratic Medium, Phys. Rev.
Lett. \textbf{74}, 5036 (1995).

\bibitem{wol1995} W. E. Torruellas, Z. Wang, L. Torner, and G. I. Stegeman,
Observation of mutual trapping and dragging of two dimensional spatial
solitary waves in a quadratic medium, Opt. Lett. \textbf{20}, 1949 (1995).

\bibitem{wprl1997} W. J. Firth and D. V. Skryabin, Optical Solitons Carrying
Orbital Angular Momentum, Phys. Rev. Lett. \textbf{79}, 2450 (1997).

\bibitem{lel1997} L. Torner and D. V. Petrov, Azimuthal instabilities and
self breaking of beams into sets of solitons in bulk second harmonic
generation, Electron. Lett. \textbf{33}, 608 (1997).

\bibitem{dol1998} D. V. Petrov, L. Torner, J. Martorell, R. Vilaseca, J. P.
Torres, and C. Cojocaru, Observation of azimuthal modulational instability
and formation of patterns of optical solitons in a quadratic nonlinear
crystal, Opt. Lett. \textbf{23}, 1444 (1998).

\bibitem{ipre2001} I. Towers, A. V. Buryak, R. A. Sammut, and B. A. Malomed,
Stable localized vortex solitons, Phys. Rev. E \textbf{63}, 055601(R) (2001).

\bibitem{dpre2004} D. Mihalache, D. Mazilu, B. A. Malomed, and F. Lederer,
Stable vortex solitons supported by competing quadratic and cubic
nonlinearities, Phys. Rev. E \textbf{69}, 066614 (2004).

\bibitem{pprl2000} P. D. Trapani, W. Chinaglia, S. Minardi, A. Piskarskas,
and G. Valiulis, Observation of Quadratic Optical Vortex Solitons, Phys.
Rev. Lett. \textbf{84}, 3843 (2000).

\bibitem{dcsf2025} D. Wu, J. Li, X. Gao, Y. Shi, Y. Zhao, L. Dong, B. A.
Malomed, N. Zhu, and S. Xu, Multicore vortex solitons in cubic-quintic
nonlinear media with a Bessel lattice potential, Chaos, Solitons \& Fractals
\textbf{192}, 116057 (2025).

\bibitem{xoe2023} X. Xu, F. Zhao, J. Huang, H. He, L. Zhang, Z. Chen, Z.
Nie, B. A. Malomed, and Y. Li, Semidiscrete optical vortex droplets in
quasi-phase-matched photonic crystals, Opt. Express \textbf{31}, 38343
(2023).

\bibitem{rol1992} R. DeSalvo, D. J. Hagan, M. Sheik-Bahae, G. Stegeman, E.
W. V. Stryland, and H. Vanherzeele, Self-focusing and self-defocusing by
cascaded second-order effects in KTP, Opt. Lett. \textbf{17}, 28 (1992).

\bibitem{cprl1995} Ch. Bosshard, R. Spreiter, M. Zgonik, and P. G\"{u}nter,
Kerr Nonlinearity via Cascaded Optical Rectification and the Linear
Electro-optic Effect, Phys. Rev. Lett. \textbf{74}, 2816 (1995).

\bibitem{soc1996} S. Trillo, A. V. Buryak, and Y. S. Kivshar, Modulational
instabilities and optical solitons due to competition of $\chi^{(2)}$ and $%
\chi^{(3)}$ nonlinearities, Optics Communications \textbf{122}, 200 (1996).

\bibitem{aol1995} A. V. Buryak, Y. S. Kivshar, and S. Trillo, Optical
solitons supported by competing nonlinearities, Opt. Lett. \textbf{20}, 1961
(1995).

\bibitem{dpre2000} D. Mihalache, D. Mazilu, L.-C. Crasovan, B. A. Malomed,
and F. Lederer, Three-dimensional spinning solitons in the cubic-quintic
nonlinear medium, Phys. Rev. E \textbf{61}, 7142 (2000).

\bibitem{lol1994} L. Torner, C. R. Menyuk, and G. I. Stegeman, Excitation of
solitons with cascaded $\chi^{(3)}$ nonlinearities, Opt. Lett. \textbf{19},
1615 (1994).

\bibitem{cpio2000} C. Etrich, F. Lederer, B. A. Malomed, T. Peschel, and U.
Peschel, Optical Solitons in Media with a Quadratic Nonlinearity, Progress
in Optics \textbf{41}, 483-568 (2000).

\bibitem{ccsf2024} C. Kong, J. Li, X. Tang, X. Li, J. Jiao, J. Cao, and H.
Deng, Composite solitary vortices of three-wave mixing in
quasi-phase-matched photonic crystals, Chaos, Solitons \& Fractals \textbf{%
187}, 115358 (2024).

\bibitem{wpla2025} W. Peng, L. Wang, L. Xiao, Y. Liao, Y. Zhao, D. Wu, and
S. Xu, Vortex solitons in a Rydberg-dressed triangular optical lattice,
Physics Letters A \textbf{534}, 130249 (2025).

\bibitem{bnp2019} B. A. Malomed, Vortex solitons: Old results and new
perspectives, Physica D: Nonlinear Phenomena \textbf{399}, 108 (2019).

\bibitem{bp2021} B. A. Malomed, Optical Solitons and Vortices in Fractional
Media: A Mini-Review of Recent Results, Photonics \textbf{8}, 9 (2021).

\bibitem{hpra2022} H. Zhang, Stabilization of higher-order vortex solitons
by means of nonlocal nonlinearity, Phys. Rev. A \textbf{105}, (2022).

\bibitem{hpra2020} H. Zhang, Z. Weng, Q. Shou, Q. Guo, and W. Hu,
Instability suppression of vector vortex solitons in nonlocal nonlinear
media, Phys. Rev. A \textbf{101}, 033842 (2020).

\bibitem{cprl2024} C. Li and Y. V. Kartashov, Stable Vortex Solitons
Sustained by Localized Gain in a Cubic Medium, Phys. Rev. Lett. \textbf{132}%
, 213802 (2024).

\bibitem{jcsf2023} J. Li and H. Zhang, Stability and adaptive evolution of
higher-order vector vortex solitons in thermally nonlinear media with
tunable transverse size, Chaos, Solitons \& Fractals \textbf{177}, 114195
(2023).

\bibitem{qol2021} Q. Shou, Z. Weng, S. Guan, H. Han, H. Huang, Q. Guo, and
W. Hu, Stable propagation of cylindrical-vector vortex solitons in strongly
nonlocal media, Opt. Lett. \textbf{46}, 2807 (2021).

\bibitem{fprl2023} F. Zhao, X. Xu, H. He, L. Zhang, Y. Zhou, Z. Chen, B. A.
Malomed, and Y. Li, Vortex Solitons in Quasi-Phase-Matched Photonic
Crystals, Phys. Rev. Lett. \textbf{130}, 157203 (2023).

\bibitem{ss1997} S. Zhu, Y. Zhu, and N. Ming, Quasi-Phase-Matched
Third-Harmonic Generation in a Quasi-Periodic Optical Superlattice, Science
\textbf{278}, 843 (1997).

\bibitem{zpre2025} Z. Fan, W. Liu, L. Wang, W. Peng, D. Wu, S. Xu, and Y.
Zhao, Vortex solitons in quasi-phase-matched photonic crystals with
competing quadratic and cubic nonlinearity, Phys. Rev. E \textbf{111},
034208 (2025).

\bibitem{jcsf2024} J. He, Y. Jiao, B. Zhou, Y. Zhao, Z. Fan, and S. Xu,
Vortex light bullets in rotating Quasi-Phase-Matched photonic crystals,
Chaos, Solitons \& Fractals \textbf{188}, 115514 (2024).

\bibitem{soe2024} S. Chen, B. Zhou, Y. Jiao, L. Wang, Y. Zhao, and S. Xu,
Vortex solitons in rotating quasi-phase-matched photonic crystals, Opt.
Express \textbf{32}, 39963 (2024).

\bibitem{ycpl2024} Y. Guo, X. Xu, Z. Chen, Y. Zhou, B. Liu, H. He, Y. Li,
and J. Xie, Three-Wave Mixing of Dipole Solitons in One-Dimensional
Quasi-Phase-Matched Nonlinear Crystals, Chinese Phys. Lett. \textbf{41},
014204 (2024).

\bibitem{tnp2018} T. Xu, K. Switkowski, X. Chen, S. Liu, K. Koynov, H. Yu,
H. Zhang, J. Wang, Y. Sheng, and W. Krolikowski, Three-dimensional nonlinear
photonic crystal in ferroelectric barium calcium titanate, Nature Photonics
\textbf{12}, 591 (2018).

\bibitem{dnp2018} D. Wei, C. Wang, H. Wang, X. Hu, D. Wei, X. Fang, Y.
Zhang, D. Wu, Y. Hu, J. Li, S. Zhu, and M. Xiao, Experimental demonstration
of a three-dimensional lithium niobate nonlinear photonic crystal, Nat.
Photonics \textbf{12}, 596 (2018).

\bibitem{snp2018} S. Keren-Zur and T. Ellenbogen, A new dimension for
nonlinear photonic crystals, Nature Photonics \textbf{12}, 575 (2018).

\bibitem{alsa2021} A. Arie, Storing and retrieving multiple images in 3D
nonlinear photonic crystals, Light Sci. Appl. \textbf{10}, 202 (2021).

\bibitem{alpr2010} A. Arie and N. Voloch, Periodic, quasi-periodic, and
random quadratic nonlinear photonic crystals, Laser \& Photonics Reviews
\textbf{4}, 355 (2010).

\bibitem{hfo2020} H. Li and B. Ma, Research development on fabrication and
optical properties of nonlinear photonic crystals, Front. Optoelectronics
\textbf{13}, 35 (2020).

\bibitem{saom2023} S. Liu, L. Wang, L. M. Mazur, K. Switkowski, B. Wang, F.
Chen, A. Arie, W. Krolikowski, and Y. Sheng, Highly Efficient 3D Nonlinear
Photonic Crystals in Ferroelectrics, Advanced Optical Materials \textbf{11},
2300021 (2023).

\bibitem{mjpd1995} M. Houe and P. D. Townsend, An introduction to method of
periodic poling for 2nd-harmonic generation, J. Phys. D \textbf{28}, 1747
(1995).

\bibitem{tol2000} T. Hatanaka, K. Nakamura, T. Taniuchi, H. Ito, Y.
Furukawa, and K. Kitamura, Quasi-phase-matched optical parametric
oscillation with periodically poled stoichiometric LiTaO3, Opt. Lett.
\textbf{25}, 651 (2000).

\bibitem{aapl2003} A. Chowdhury, H. M. Ng, M. Bhardwaj, and N. G. Weimann,
Second-harmonic generation in periodically poled GaN, Appl. Phys. Lett.
\textbf{83}, 1077 (2003).

\bibitem{jol2004} J. P. Torres, A. Alexandrescu, S. Carrasco, and L. Torner,
Quasi-phase-matching engineering for spatial control of entangled two-photon
states, Opt. Lett. \textbf{29}, 376 (2004).

\bibitem{aoe2018} A. Karnieli and A. Arie, Fully controllable adiabatic
geometric phase in nonlinear optics, Opt. Express \textbf{26}, 4920 (2018).

\bibitem{afp2021} A. Karnieli, Y. Li, and A. Arie, The geometric phase in
nonlinear frequency conversion, Front. Phys. \textbf{17}, 12301 (2021).

\bibitem{cjosab2013} C. R. Phillips, C. Langrock, D. Chang, Y. W. Lin, L.
Gallmann, and M. M. Fejer, Apodization of chirped quasi-phase-matching
devices, J. Opt. Soc. Am. B \textbf{30}, 1551 (2013).

\bibitem{foe2021} F. Zhao, J. L\"{u}, H. He, Y. Zhou, S. Fu, and Y. Li, Geometric phase with full-wedge and
half-wedge rotation in nonlinear frequency conversion, Opt. Exp. \textbf{29}%
, 21820 (2021).

\bibitem{ypra2020} Y. Li, Adiabatic geometric phase in fully nonlinear
three-wave mixing, Phys. Rev. A \textbf{3}, 101 (2020).

\bibitem{gjosab2000} G. G. Luther, M. S. Alber, J. E. Marsden, and J. M.
Robbins, Geometric analysis of optical frequency conversion and its control
in quadratic nonlinear media, J. Opt. Soc. Am. B \textbf{17}, 932 (2000).

\bibitem{jpla2021} J. L\"{u}, F. Zhao, W. Pang, and Y. Li, Constant
adiabatic geometric phase in three-wave mixing under different depletion
levels, Phys. Lett. A \textbf{397}, 127266 (2021).

\bibitem{gjosab2013} G. Porat and A. Arie, Efficient, broadband, and robust
frequency conversion by fully nonlinear adiabatic three wave mixing, J. Opt.
Soc. Am. B \textbf{30}, 1342 (2013).

\bibitem{fylpra2008} F. Ye, Y. V. Kartashov, and L. Torner, Stabilization of dipole solitons in nonlocal nonlinear media, Phys. Rev. A \textbf{77}, 043821 (2008).

\bibitem{sodpre2006} S. Skupin, O. Bang, D. Edmundson, and W. Krolikowski, Stability of two-dimensional spatial solitons in nonlocal nonlinear media, Phys. Rev. E 73, 066603 (2006).

\bibitem{bel2003} B. B. Baizakov, B.A. Malomed, and M. Salerno,
Multidimensional solitons in periodic potentials, Europhys. Lett. \textbf{63}%
, 642 (2003).

\bibitem{jol2003} J. Yang and Z. H. Musslimani, Fundamental and vortex
solitons in a two-dimensional optical lattice, Opt. Lett. \textbf{28}, 2094
(2003).

\bibitem{aprl2005} A. Ferrando, Vorticity Cutoff in Nonlinear Photonic
Crystals, Phys. Rev. Lett. \textbf{95}, 043901 (2005).

\bibitem{kprl2017} K. Krupa, K. Nithyanandan, U. Andral, P. Tchofo-Dinda,
and P. Grelu, Real-Time Observation of Internal Motion within Ultrafast
Dissipative Optical Soliton Molecules, Phys. Rev. Lett. \textbf{118}, 243901
(2017).

\bibitem{jprl2003} J. W. Fleischer, T. Carmon, M. Segev, N. K. Efremidis,
and D. N. Christodoulides, Observation of Discrete Solitons in Optically
Induced Real Time Waveguide Arrays, Phys. Rev. Lett. \textbf{90}, 023902
(2003).

\bibitem{nqe1973} N. G. Vakhitov and and A. A. Kolokolov, Stationary
solutions of the wave equation in a medium with nonlinearity saturation,
Radiophys. Quantum Electron. \textbf{16}, 783 (1973).

\bibitem{http} $27\,\text{pm/V}$ is the largest value of the lithium
niobate's nonlinear tensor, $d_{33}$. Generally, $d_0 = (2/\pi)d_{33}$. In
Eq.~(5), the coefficient of $(2/\pi)$ is separated, hence we set $d_0 =
d_{33}$ in this Letter; \url{http://en.wikipedia.org/wiki/Lithium_niobate}.

\bibitem{djosab1997} D. E. Zelmon, D. L. Small, and D. Jundt, Infrared
corrected Sellmeier coefficients for congruently grown lithium niobate and 5
mol. \% magnesium oxide-doped lithium niobate, J. Opt. Soc. Am. B \textbf{14}%
, 3319 (1997).

\bibitem{ispinjose} I. A. Kulagin, R. A. Ganeev, V. A. Kim, A. I.
Ryasnyansky, R. I. Tugushev, T. Usmanov, and A. V. Zinoviev, Nonlinear
refractive indices and third-order susceptibilities of nonlinear-optical
crystals, in Proceedings of the SPIE 4972, Nonlinear Frequency Generation
and Conversion: Materials, Devices, and Applications II (SPIE, San Jose, CA,
2003).
\end{thebibliography}
\end{document}